\documentclass[iop, apj, final]{emulateapj}
\usepackage[stretch=10]{microtype}
\usepackage{apjfonts}
\usepackage{graphicx}
\usepackage{amsmath}
\usepackage{amssymb}
\usepackage{booktabs}
\usepackage[svgnames,hyperref]{xcolor}
\usepackage[pagebackref=false]{hyperref}
\usepackage[all]{hypcap}

\hypersetup{
	colorlinks=true,
	linkcolor=NavyBlue,
	citecolor=NavyBlue,
	urlcolor=NavyBlue}

\newcommand{\mchirp}{\mathcal{M}}
\newcommand{\mbh}{M_\text{BH}}
\newcommand*{\negthick}{%
  \mkern-\thickmuskip
}

\def\equationautorefname~#1\null{Equation~(#1)\null}

\shorttitle{Cosmic variance in the nHz GW background}
\shortauthors{Roebber et al.}

\begin{document}

\title{Cosmic variance in the nanohertz gravitational wave background}

\author{Elinore Roebber\altaffilmark{*} and Gilbert Holder}
\email[* Email: ]{roebbere@physics.mcgill.ca}
\affil{Department of Physics, McGill University, Montr\'eal, QC,
Canada H3A 2T8}

\author{Daniel E. Holz}
\affiliation{Enrico Fermi Institute, Department of Physics, and Kavli Institute
for Cosmological Physics\\University of Chicago, Chicago, IL 60637}

\and

\author{Michael Warren}
\affil{Theoretical Division, LANL, Los Alamos, NM 87545}

\begin{abstract}
We use large N-body simulations and empirical scaling relations between dark matter halos, galaxies, 
and supermassive black holes to estimate the formation rates of supermassive black hole binaries and 
the resulting low-frequency stochastic gravitational wave background (\textsc{gwb}).  We find this 
\textsc{gwb} to be relatively insensitive ($\lesssim10\%$) to cosmological parameters, with only slight 
variation between \textsc{wmap5} and Planck cosmologies.  We find that uncertainty in the astrophysical 
scaling relations changes the amplitude of the \textsc{gwb} by a factor of $\sim 2$.  Current observational 
limits are already constraining this predicted range of models.  We investigate the Poisson variance 
in the amplitude of the \textsc{gwb} for randomly-generated populations of supermassive black holes, 
finding a scatter of order unity per frequency bin below $10$~nHz, and increasing to a factor of $\sim 10$ 
near $100$~nHz.  This variance is a result of the rarity of the most massive binaries, which dominate the 
signal, and acts as a fundamental uncertainty on the amplitude of the underlying power law spectrum.  
This Poisson uncertainty dominates at $\gtrsim 20$~nHz, while at lower frequencies the dominant 
uncertainty is related to our poor understanding of the astrophysical scaling relations, although very low 
frequencies may be dominated by uncertainties related to the final parsec problem and the processes 
which drive binaries to the gravitational wave dominated regime. Cosmological effects are negligible at 
all frequencies.
\end{abstract}

\keywords{black hole physics --- gravitational waves --- large-scale structure of universe}

\section{Introduction}

Supermassive black holes, with masses $\gtrsim 10^6 M_\odot$, appear to 
reside at the center of nearly every moderate to massive galaxy \citep{kormendy95}.  Mergers 
between galaxies are relatively common \citep{lotz11}, suggesting that the formation of binary 
supermassive black holes (\textsc{smbbh}s) should also occur regularly \citep{begelman80}. Such 
\textsc{smbbh}s should produce considerable gravitational radiation in a frequency band of 
$\sim10^{-9}$--$10^{-7}$ Hz  \citep[e.g.][]{thorne87, maggiore08}. Careful monitoring of arrival times 
of radio pulses from a large number of individual pulsars could allow detection of this gravitational 
radiation.  This idea is implemented in  pulsar timing arrays (\textsc{pta}s) \citep{hellings1983, 
foster90, shannon15, arzoumanian15, lentati15}.  Recent estimates  \citep[e.g.][]{sesana13a, 
mcwilliams14, ravi15a, rosado15a} suggest that detection of gravitational radiation with 
\textsc{pta}s may occur in the near future.

The gravitational radiation due to \textsc{smbbh}s is expected to form an approximately isotropic 
stochastic background, with a spectrum at relatively low frequencies set by the well-known rate 
at which the binary's orbit decays due to emission of gravitational radiation \citep{phinney01}.  
The amplitude is set by the total population of emitting \textsc{smbbh}s, which can be 
estimated from the current understanding of galaxy merger rates and the apparent ubiquity 
of supermassive black holes in galaxies of at least moderate size.

The amplitude of gravitational radiation produced by \textsc{smbbh}s also depends on a large 
number of environmentally dependent and often poorly-understood astrophysical variables: 
the characteristic masses of black holes in the centers of galaxies \citep{lauer07, kormendy13}; 
merger rates of the relevant galaxies; dynamical friction \citep{chandrasekhar43, boylan-kolchin08};  
the physical mechanisms removing sufficient energy and angular momentum from the binary 
system \citep[e.g.][]{kocsis11, ravi14a} to bring the black holes close enough to eventually merge 
through the emission of gravitational radiation \citep[the `final parsec problem,' see][]{merritt05}; 
the eccentricity of the binary black hole orbits \citep{peters63}; and the evolution of all of these 
effects as a function of cosmic time.

There have been many predictions of the expected gravitational wave background (hereafter 
\textsc{gwb}) produced by \textsc{smbbh}s, with a variety of approaches used: semi-analytic 
predictions with merger rates based on the Press-Schechter formalism \citep{wyithe03, enoki04, 
enoki07, sesana04, sesana08, rosado11}; N-body simulations dressed with black holes using 
prescriptions for how various types of galaxies form \citep{sesana09, kocsis11, kulier15a, ravi14a}; 
and empirically-based estimates of the galaxy merger rate from observed numbers of close pairs 
\citep{jaffe03, sesana13a, ravi15a}.  In all cases, uncertainties in how black holes populate galaxies 
lead to a large theoretical uncertainty in the expected \textsc{gwb} level.

In this paper we use very large cosmological N-body simulations (dark matter only) with up-to-date 
cosmological parameters and recent estimates of various scaling relations to relate dark matter halo 
mass with black hole mass and calculate the expected \textsc{gwb}.  We will briefly explore the 
uncertainty in these astrophysical scaling relations, but will primarily focus on the fundamental scatter 
in the \textsc{gwb} due to Poisson statistics in the population of \textsc{smbbh}s.

The organization of the paper is as follows:  In \autoref{sec:calc_gwb} we describe our procedure for 
calculating the \textsc{gwb} starting with merger trees from N-body simulations.  We explain our 
prescription for placing galaxies within dark matter halos and black holes within galaxies, as well 
as our Monte Carlo sampling of the population of \textsc{smbbh}s and subsequent calculation of 
the \textsc{gwb}.  In \autoref{sec:discussion} we discuss the characteristics of our predicted 
signal.  This includes the amplitude of the signal, differences between simulations, exploration of 
the effects of astrophysical uncertainties, as well as the scatter in the amplitude as a function of 
frequency for many different Monte Carlo realizations.  \autoref{sec:conclusion} presents our 
conclusions and discussion of future directions.

\section{Calculating the Gravitational Wave Background}
\label{sec:calc_gwb}

In this section, we explain how we generate a representative population of \textsc{smbbh}s beginning 
with dark matter merger trees from N-body simulations, and calculate the \textsc{gwb} produced by 
this population in the \textsc{pta} frequency band.  In the subsections below we discuss the dark matter 
simulations used, describe how we populate dark matter halos with black holes, explain our prescription 
for calculating binary black hole formation rates from halo merger trees, review the calculation of 
gravitational wave strain from \textsc{smbbh}s, and finally put all these pieces together to describe 
how we produce individual realizations of the population of \textsc{smbbh}s and calculate the resulting 
\textsc{gwb}.

\subsection{N-body Simulations}

\begin{table}
\centering
\begin{tabular}{lcc}
	\toprule
	                                     			&        Dark Sky                 &      MultiDark          \\ 
	\midrule
	$\Omega_m$                			&          0.32                       &      0.27                     	\\
	$\Omega_\Lambda$     			&          0.68                       &      0.73                     	\\
	$h$                                			&          0.67                       &      0.7                      	\\
	$\sigma_8$                   			&          0.83                       &      0.82                     	\\
	Box volume $(\text{Mpc}^3/h^3)$  	& 	   $1000^3$     	      	&      $1000^3$  		\\
	Particle mass  ($M_\odot/h$)    		&         $10^{10}$ 	      	&   	$9\times 10^9$		\\
	Number of particles				&	  $2048^3$	 	&	$2048^3$			\\
	Number of snapshots   			&           13                         &     43                     		\\ 
	Longest timestep (Myr)  			&	    1600		      	&	500				\\
	Shortest timestep (Myr) 			&	    470		      	& 	110				\\
	Lowest snapshot redshift           	&            0                         &       0                         	\\
	Highest snapshot redshift          	&            4                         &       10                       	\\
	\bottomrule
\end{tabular}
\caption{Properties of the simulations and their merger trees }
\label{tab:simprops}
\end{table}

We begin our calculation with merger trees from two recent large-scale dark matter simulations.
Their properties are summarized in \autoref{tab:simprops}.  The first is the Dark Sky simulation 
\citep{warren13a}, which uses a set of cosmological parameters based on Planck \citep{planckxvi13}. 
The second is MultiDark, or Big Bolshoi \citep{riebe11, prada12}, which uses the same initial 
conditions as the Bolshoi simulation \citep{klypin11a}, and is consistent with \textsc{wmap}5 
parameters \citep{komatsu09}, and remains consistent with \textsc{wmap}9 \citep{hinshaw13}.  
The \textsc{wmap}5 and Planck cosmologies are similar, particularly for the value of $\sigma_8$, 
which is important for structure formation.  Differences remain, especially for the parameters 
$\Omega_m$ and $h$.  Both of these cosmologies are significantly different from that chosen 
for the Millennium simulation, which is consistent with the \textsc{wmap}1 cosmology
\citep{springel05}.  

We use merger trees produced from these simulations by the halo finder \textsc{Rockstar} 
\citep{behroozi13a} and merger tree code Consistent Trees \citep{behroozi13b}.  Throughout this
work, we will use the term ``halo'' to mean any subhalo or host halo listed in the merger trees.

Halo mass functions for Dark Sky and MultiDark at $z=0$ and $z=1$ are shown in the top left panel 
of \autoref{fig:all_mfuns}.  The two simulations have similar volumes and mass resolutions.  
Both mass functions are essentially complete between $\sim 10^{12} M_\odot$ and 
$\sim 10^{15} M_\odot$ at low redshift.  The slight offset in the amplitudes of the halo mass 
functions is a result of their different cosmologies.  In addition to the cosmology, the two simulations 
differ in their time resolution and the redshift range in which the snapshots are taken.  We will 
consider the effects of these differences throughout our calculation.

\subsection{Populating Halos with Black Holes}
\label{sec:place_bhs}

Supermassive black holes are correlated with properties of their host galaxies \citep[for a review, see][]{kormendy13}.  
In order to make predictions about the population of supermassive black holes, we first need to characterize the 
population of galaxies inhabiting the dark matter halos in each simulation.  We use the scaling relation 
between halo mass and galaxy stellar mass calculated by \citet{behroozi13c} to derive the stellar masses 
of galaxies inside the halos.   
 
This relation was derived by requiring consistency between abundance matching calibrated to 
observed galaxy stellar mass functions and calculations of the star formation rate.  They fit the 
stellar mass--halo mass relation to a 5-parameter model which behaves as a power law at low 
mass and a subpower law at high mass and varies as a function of redshift between $0<z<8$. 
This form was chosen to provide a better fit to the observed stellar mass function than the 
commonly-used double power law \citep{behroozi13c}.  Their relation also contains an evolving 
intrinsic scatter in the stellar mass at low redshifts.

\begin{figure*}
	\includegraphics{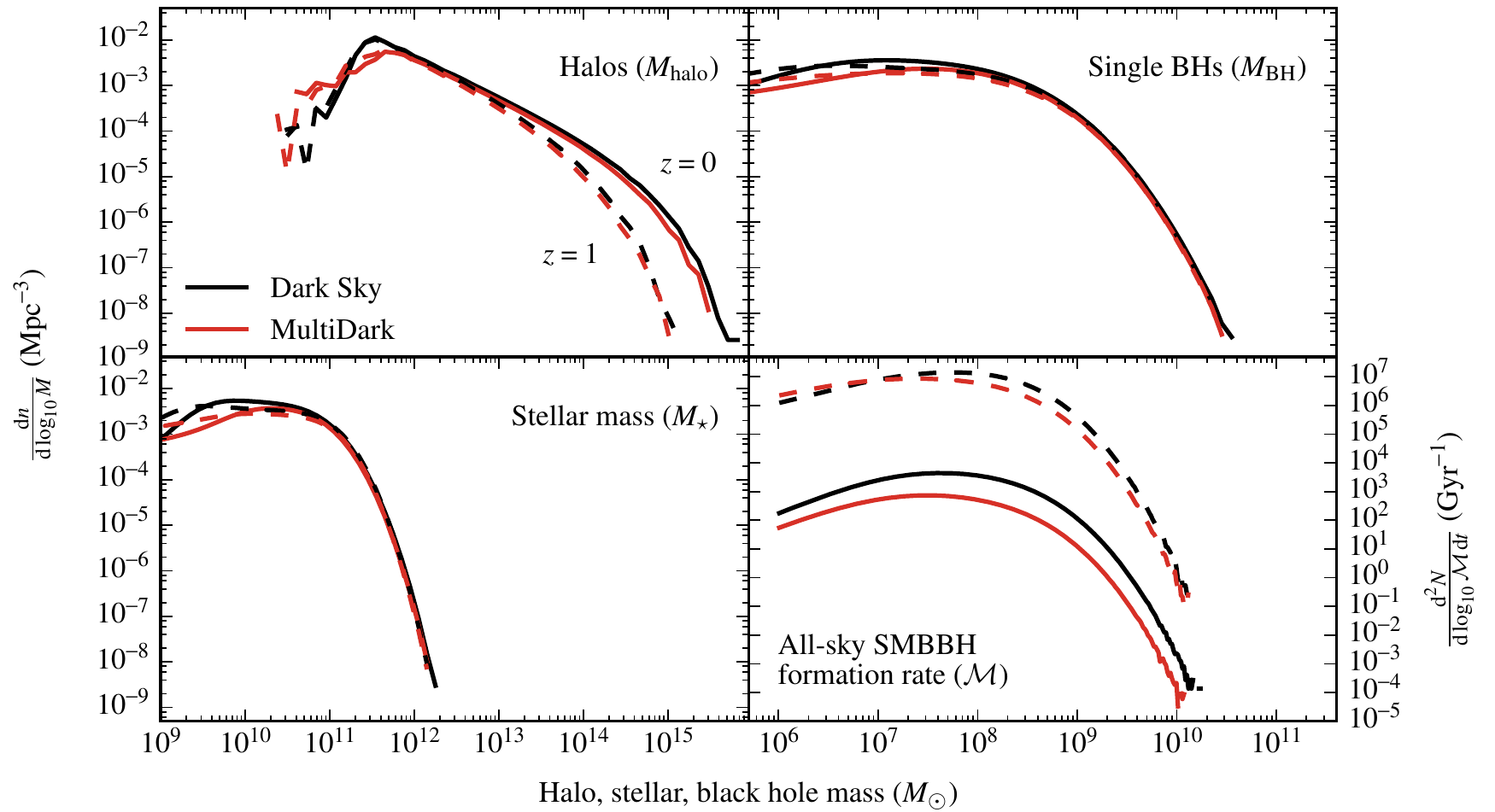}
	\caption{Halo, stellar, black hole, and \textsc{smbbh} mass functions
			calculated for the Dark Sky and MultiDark merger trees as described in the text.  
			Solid lines show the mass functions at $z=0$ and dashed lines represent $z=1$, 
			except for binaries, for which solid lines represent an interval of 
			$0\lesssim z \lesssim 0.05$ and dashed lines represent $1 \lesssim z \lesssim 1.5$.  
			The resolution limit can be clearly seen in the peaks of the halo mass functions 
			between $10^{11} M_\odot$ and $10^{12} M_\odot$.  This limit is then propagated 
			with scatter to the galaxy, single black hole, and \textsc{smbbh} mass functions.  
			The \textsc{smbbh} formation rates represent the number of binaries produced 
			per Gyr, integrated over the volume in each redshift bin.  The shift in amplitude 
			between the two redshift intervals is primarily due to the much greater volume at 
			$1 < z < 1.5$ than at $0 < z < 0.05$. 
			}
	\label{fig:all_mfuns}
\end{figure*}

We use this relation to populate halos in our two simulations with galaxies.  The 
intrinsic scatter is accounted for by randomly drawing from a normal distribution with the appropriate 
standard deviation.  The resulting stellar mass functions are shown in the 
bottom left panel of \autoref{fig:all_mfuns}.   The evolution of the stellar mass function at masses 
above $10^{11} M_\odot$ is not significant at redshifts $0<z<1$, which will turn out to be the redshift 
range of most interest for our results.  Below $10^{11}M_\odot$, the stellar mass function begins to 
drop due to the minimum halo mass resolved in the simulations.  A small offset between the two 
simulations remains, but is less significant than in the halo mass functions.  

We model different populations of galaxies by splitting galaxies between `quiescent' and 
`star-forming' populations based on the evolving mass functions from \citet{moustakas13}.  
After this split, we calculate bulge masses for each population based on the relations shown 
in Figure~1 of \citet{lang14}.  Our model does not take into account environmental dependence 
or merger history, but is calculated for the population as a whole.

With this calculation of the galaxy bulge mass, we can determine the distribution of black holes. 
We place black holes inside the galaxies using the \citet{kormendy13} stellar bulge mass--black hole 
mass scaling relation, randomly drawing the black hole masses according to the intrinsic scatter:
\begin{equation}
	\left( \frac{\mbh}{10^9 M_\odot} \right)= 0.049 \left( \frac{M_\star}{10^{11} M_\odot} \right)^{1.16} 
		\negthick \negthick,  \quad  \text{scatter} = 0.29 \text{ dex}.
\end{equation}
This relation incorporates recent measurements and recalculations of black hole masses, 
especially for the largest black holes present in brightest cluster galaxies, which may have been 
systematically underestimated previously \citep[e.g.][]{gebhardt11, hlavacek-larrondo12, 
mcconnell12, rusli13}.  Additionally, they restrict the sample of black holes to those with host 
galaxies that are ellipticals and spirals with so-called classical bulges in order to form a clean 
sample of ``objects that are sufficiently similar in formation and structure'' \citep{kormendy13}.    

We do not apply any explicit prescriptions for black hole mass growth through accretion (or 
mergers).  Instead, we assume that this scaling relation applies at all redshifts, implicitly requiring 
that black holes grow along with their host galaxies.  For merging halos, the scaling relations 
are applied at the timestep before merger.  The resulting black hole mass functions are shown 
in the top right panel of \autoref{fig:all_mfuns}.  The mass functions are remarkably consistent 
for $\mbh \gtrsim 10^9 M_\odot$ for $0 < z < 1$.  For $\mbh \lesssim 10^8 M_\odot$, the 
differences between simulations are once more due to the mass resolution of the simulations. 
The point at which the two simulations deviate shows where the mass functions become untrustworthy.  
This mass changes somewhat with redshift, but remains of order $10^8 M_\odot$.  

In addition to the uncertainty in our mass functions due to the simulation resolution, we expect additional 
error due to our choice of scaling relations.  
In particular, the relationship between stellar mass and bulge mass, which is related to galaxy morphology, 
is not well known.  Moreover, the distribution of ellipticals versus spirals depends strongly on environment 
through the well-studied morphology--density relation \citep{dressler80}.  Accurate modeling of the 
merger rate for each galactic morphological type would require detailed semi-analytic models, so 
we treat the relation between total stellar mass and bulge mass as an astrophysical uncertainty 
in our calculations.

The uncertainty in the $M_\text{bulge}$--$\mbh$ relations increases at both high
and low black hole mass.  For the latter case, there is a significant population of black holes 
which no longer be fit tight scaling relations \citep{kormendy13}.  
As a result, our scaling relations are inaccurate for $\mbh \lesssim 10^8 M_\odot$ and for 
$\mbh \gtrsim 10^{10} M_\odot,$ but the effect of this on the signal is likely to be smaller than 
that due to the mass resolution.

\subsection{Black Hole Binaries Form When Halos Merge}
\label{sec:find_bbh}

We use the supermassive black hole mass functions calculated in the previous section to assign \textsc{smbbh}s
to merging halos identified by the merger trees.  Consistent Trees \citep{behroozi13b} builds merger 
trees by tracking halos across multiple timesteps.  It uses this history to determine when a halo 
(typically a subhalo) has merged into a larger halo (the halo creating the strongest tidal field).  We 
use the halo masses calculated at the timestep before merger to calculate black hole masses.

For each halo merger we calculate the masses of the black holes residing in the two largest dark 
matter halo progenitors, and assign an equivalent binary to the descendant halo.  We remove minor 
mergers from our list by performing a cut in the stellar mass ratio of progenitors such that 
$M_{\star, 2}/M_{\star, 1} \geq 0.05$, where $M_{\star, 1}$ is the progenitor with the larger stellar 
mass and $M_{\star, 2}$ is the progenitor with the smaller stellar mass.  We assume that dynamical 
friction proceeds quickly \citep[see][]{dotti07} and that the final parsec problem is solved in all cases, 
so that all black holes form a binary as soon as the progenitor halos merge.  We do not consider 
any accretion onto either black hole during the galaxy merger or the possibility of triple systems.  

With these assumptions, we can calculate the mass function of \textsc{smbbh}s formed between 
snapshots for each simulation.  We parameterize the mass of black hole binaries using the chirp 
mass:
\begin{equation}
	\mchirp \equiv \mu^{3/5} M_\text{tot}^{2/5} = \frac{(m_1m_2)^{3/5}}{(m_1 + m_2)^{1/5}}
	\label{eq:mchirp}
\end{equation}
and the black hole mass ratio 
\begin{equation}
	q = \frac{m_2}{m_1}, \quad m_1 > m_2,
	\label{eq:q}
\end{equation}
where $m_1$ and $m_2$ are the masses of the black holes in the binary.  
We allow a mass range of $10^6 M_\odot < \mchirp < 10^{10.5} M_\odot$ and a 
black hole mass ratio range of $5\times 10^{-4} < q < 1$ to ensure that all binaries producing 
significant signal are included. The mass function is logarithmically binned in both $\mchirp$ 
and $q$, with bin sizes chosen so that bins are always less than $10\%$ of each quantity. 

In order to ensure a smooth distribution which accounts for the intrinsic scatter in the scaling 
relations, we repopulate the merging halos in each simulation 10,000 times and use the average 
to calculate the final mass function of \textsc{smbbh}s.  Although we are able to account for the 
scatter in the scaling relations this way, we only have a single set of merger trees for each 
simulation, and so cannot similarly calculate the scatter in the merger rates of halos.  We do 
not expect this to significantly affect our results, since most halos hosting black holes of interest 
are well-represented in the simulation boxes.  

The rate at which \textsc{smbbh}s of different masses are produced is shown in the bottom right 
panel of \autoref{fig:all_mfuns}.  
Rates are derived by counting the number of mergers per chirp mass bin, 
integrating over volume, and dividing by the time elapsed 
for each redshift range.  We show the intervals  $0 \lesssim z \lesssim 0.05$ and 
$1 \lesssim z \lesssim 1.5$, which correspond to single timesteps for Dark Sky and the sum of 
multiple timesteps for MultiDark.  
Dark Sky has a slightly higher binary black hole production rate than MultiDark at low redshift, 
which largely disappears by $z \sim 1$.  
This is due to a higher halo merger rate at low redshifts. 

\subsection{Gravitational Radiation from Black Hole Binaries} 
\label{sec:gw_sig}

At separations $\ll 1$~pc, the evolution of \textsc{smbbh}s becomes dominated by emission 
of gravitational radiation.  As it emits gravitational waves, the radius of a binary will shrink and the 
orbit will circularize.  For a binary in a circular orbit, the gravitational radiation emitted will be at a 
single frequency $f_\text{emit} = 2f_\text{orbit}$.  If the binary is at cosmological distances, the 
frequency seen by an observer will be redshifted such that 
\begin{equation}
	f_\text{obs} = \frac{f_\text{emit}}{1+z} = \frac{2f_\text{orbit}}{1+z}.
\end{equation}
As it emits, the binary's orbit will decay and the frequency will increase as 
\citep[e.g.][]{thorne87, cutler94, wyithe03, sesana08}
\begin{equation}
	\frac{df_\text{emit}}{dt} = \frac{96}{5} \frac{(G \mchirp)^{5/3}}{\pi c^5} (\pi f_\text{emit})^{11/3},
	\label{freq_decay}
\end{equation}
or equivalently, the relationship between the emitted frequency at a time $t$ after the binary begins 
emitting at frequency $f_0$ can be written as
\begin{equation}
	(\pi f_0)^{-8/3} - (\pi f_\text{emit})^{-8/3} = \frac{256}{5} \frac{(G \mchirp)^{5/3}}{c^5} \, t.
	\label{freq(t)1}
\end{equation}

We consider binaries in circular orbits emitting gravitational waves at frequencies between the 
initial frequency where gravitational radiation becomes the dominant mechanism bringing the 
black holes together, and a final frequency just before coalescence.

We choose our initial frequency based on the \citet{quinlan96} estimate for when gravitational 
radiation becomes efficient:
\begin{equation}
		f_\text{0} = 2.2 \text{ nHz} \left(\frac{m_1 + m_2 }{2\times10^8 M_{\odot} }\right)^{1/5}
						\left( \frac{(10^8 M_{\odot})^{2}}{m_1m_2} \right)^{3/10} \negthick.
		\label{f0}
\end{equation}
Although the exact frequency where gravitational radiation dominates over other processes 
bringing the binary together depends on how the final parsec problem is solved, changing our 
initial frequency does not significantly affect our results, provided it is neither well inside the visible 
frequency range, nor sufficiently low that most binaries will not merge within a
Hubble time. 
Both of these situations could be reproduced by astrophysical processes related to the final 
parsec problem.  The first might be caused by high eccentricities or environmental effects which 
dominate even when the binaries emit significant gravitational radiation \citep[e.g.][]{enoki07, 
kocsis11, ravi14a}.  The second would occur if the final parsec problem is not solved for all 
binaries \citep[e.g.][]{mcwilliams14}.

Following \citet{hughes02}, we estimate the final frequency from the innermost stable circular orbit 
for a Schwarzschild black hole of equivalent mass:
\begin{align}
	f_\text{\textsc{isco}} = 22 \,\mu \text{Hz} \left(\frac{2\times10^8 M_\odot}{m_1 + m_2}\right).
	\label{fisco}
\end{align}
Since the orbit of the binary decays extremely rapidly towards the end of its lifetime, the exact value 
chosen for this frequency does not significantly affect the total lifetime of the binary nor the range of 
frequencies over which it is observed.

Using \autoref{freq(t)1} and the above initial and final frequencies, we can write the emitted 
frequency as a function of time more simply:
\begin{align}
	f_\text{emit}(t) = \frac{1}{\pi} \left[  \left(1 - \frac{t}{T}\right) (\pi f_0)^{-8/3} + 
				\frac{t}{T} \left(\pi f_\text{\textsc{isco}}\right)^{-8/3} \right]^{-3/8}
	\label{freq(t)}
\end{align}
where $T$ is the lifetime of the binary, namely the time that passes between when the binary 
begins emitting at $f_0$ and when it reaches $f_\text{\textsc{isco}}$.  We will use this relation in the 
next section to randomly sample the frequency distribution of gravitational wave sources.  

The amplitude of gravitational waves produced by a single binary at this characteristic frequency, 
averaged over angle and polarization, is given by \citep[e.g.][]{thorne87}:
\begin{equation}
	h(f_\text{emit}) = \sqrt{\frac{32}{5}} \frac{(G\mchirp)^{5/3}}{c^4 R(z)} (\pi f_\text{emit})^{2/3},
	\label{h1binary}
\end{equation}
where $R(z)$ is the comoving distance to the \textsc{smbbh}.  

\subsection{The GWB Produced by a Population of Binaries}
\label{sec:gwb_from_pop}

The \textsc{gwb} is the result of an incoherent sum of many gravitational wave signals and depends 
on the properties of the population which produces it.  We wish to investigate the \textsc{gwb} from 
a statistical point of view, so we will use Monte Carlo selection to generate many different 
realizations of the population of binaries observed to be emitting gravitational radiation.

In particular, for each redshift interval $\Delta z_i$, we will convert the comoving number density 
of binaries produced within the simulation box, $n(\mchirp, q, \Delta z_i)$, into the distribution 
$N^\text{obs}_i (\mchirp, q)$ representing the population of \textsc{smbbh}s seen by an observer 
in a universe with the same underlying distributions as the simulation.  We will draw from this 
distribution to simulate the population of \textsc{smbbh}s out to $z\sim 4$ in an individual 
realization of the universe.  This process is analogous to that done by e.g.\ \citet{sesana08}.

To begin, we calculate the comoving number density of binaries with $\mchirp \ge 10^6 M_\odot$ 
formed between each snapshot in the simulation, as described in \autoref{sec:find_bbh}. 
We write the comoving number density as $n(\mchirp, q, \Delta z_i)$. 
We then convert this to the expected number of binaries formed within the volume of the shell 
corresponding to the redshift interval $\Delta z_i$:
\begin{equation}
	N_i (\mchirp, q) = n(\mchirp, q, \Delta z_i) V_c(\Delta z_i).
\end{equation}

Next, we account for the fact that binaries of different masses have different lifespans, and therefore 
are not equally likely to be observed.  As a result of the mass dependence in 
\autoref{freq_decay}, heavier binaries merge much more quickly than light ones, and so 
any observation of the \textsc{smbbh} population will tend to see a smaller percentage of heavy 
binaries than would naively be expected from the bottom right panel of \autoref{fig:all_mfuns}. 

We use this property to transform the distribution of binaries formed within a redshift shell to the 
distribution of binaries seen to be emitting gravitational radiation by an observer at $z=0$:
\begin{equation}
	N^\text{obs}_i (\mchirp, q) = N_i (\mchirp, q)  \times \text{min}\left(\frac{ T(\mchirp, q)}{\Delta t_i}, 1\right), 
\end{equation}
where $T(\mchirp, q)$ is the lifetime of binaries in the bin and $\Delta t_i$ is the time between 
snapshots.   This distribution represents the number of sources present on our lightcone.  We 
will draw from this population of \textsc{smbbh}s to produce the \textsc{gwb}.  

For each redshift interval $z_i$, we Poisson sample the distribution $N^\text{obs}_i (\mchirp, q)$ 
to produce a single realization of the population.  This is necessary since many bins of 
$N^\text{obs}_i (\mchirp, q)$ have less than one source on average, particularly at low redshifts 
and high masses where the resulting gravitational signal is brightest.  

Having drawn the masses of the binaries in our population, we select corresponding redshifts and 
frequencies at which the binaries emit.  Since all binaries are already on our lightcone, we choose 
frequencies by drawing randomly from within each binary's lifetime, and then converting that time to 
a frequency according to \autoref{freq(t)}.  We choose redshifts by drawing from a 
volume-weighted distance distribution corresponding to the redshift interval, and then converting 
the distance to a redshift.  This is important at low redshifts in order to avoid spurious sources at 
redshifts very close to zero.  We calculate the properties of the population of binaries anew for each 
timestep, neglecting any binaries which might still exist in the next, since most timesteps are longer 
than the lifetimes of the binaries producing most of the signal.

With this population of observed binaries we can calculate the total strain in the \textsc{gwb} for a single 
realization of the universe.  To allow comparison with \textsc{pta} upper limits, we bin the signal by the 
inverse of the total observation time.  The minimum frequency is likewise set by the total observation 
time and the maximum frequency is set by the cadence.  To illustrate, we will assume a \textsc{pta} 
observation time of 25 years with observations occurring every six weeks.  The frequency bins 
therefore have widths
\begin{equation}
	\Delta f = \frac{1}{T_\text{obs}} = (\text{25 years})^{-1} \approx 1.3 \text{ nHz},
\end{equation}
and are located between $f_\text{min} \approx 1.3$~nHz and
\begin{equation}
	f_\text{max} = \frac{1}{2} (\text{6 weeks})^{-1} \approx 100 \text{ nHz}.
\end{equation}
We will test the effect of these assumptions in \autoref{sec:vary_bin_widths}.

\begin{figure}
	\includegraphics{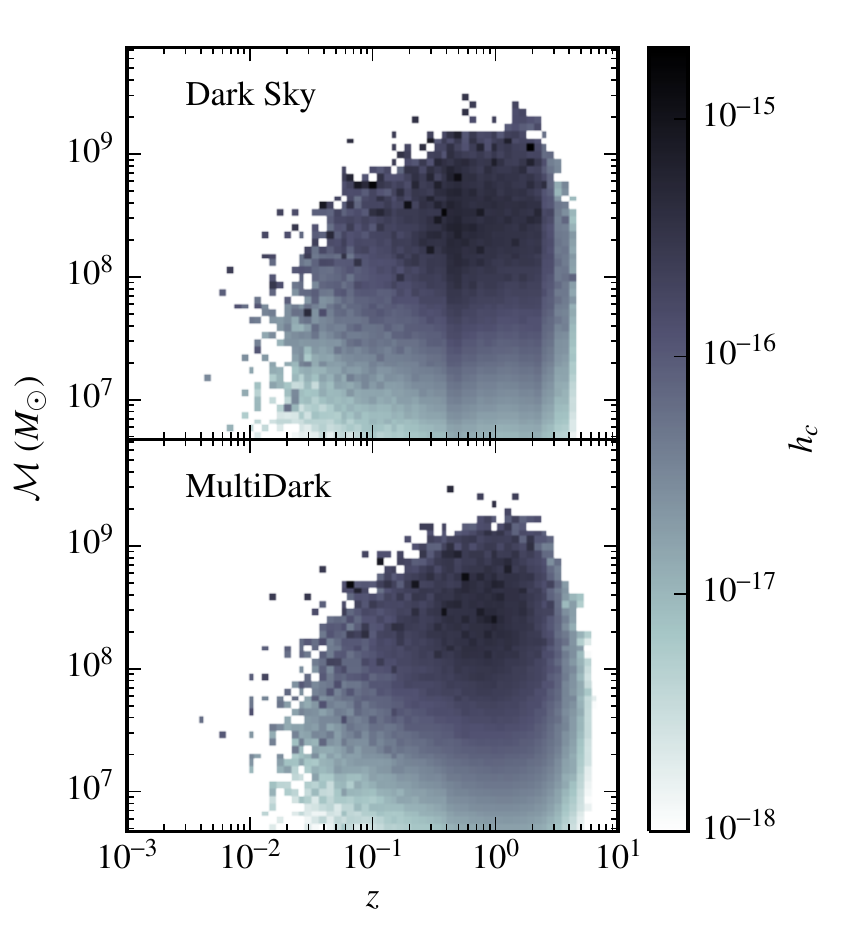}
	\caption{A single Monte Carlo realization for each simulation of the total strain produced by  
			a population of \textsc{smbbh}s as a function of chirp mass and redshift.
			We include all binaries with frequencies between $f_\text{min}$ and 
			$f_\text{max}$.  Varying strain amplitude between adjacent bins is mostly due 
			to sources emitting gravitational waves at different frequencies. }
	\label{fig:hc_of_M_z}
\end{figure}

We calculate the gravitational radiation produced by each binary according to \autoref{h1binary}.  
The characteristic strain $h_c$ in each bin $ f_n$ is given by a weighted sum over the squared strain of 
all sources observed at frequencies $f_n < f_i^\text{obs} < f_n + \Delta f$ \citep{thorne87}: 
\begin{equation} 
	h_c ( f_n) =  \left(\sum_i h_i^2 \! \left(f_i^\text{emit}\right) f_i^\text{obs} t_i^\text{bin} \right)^{1/2},
	\label{eq:GWB}
\end{equation}
where $t_i^\text{bin} \le T_\text{obs}$ is the total time that the binary is observed at a frequency 
within bin $f_n$.  The vast majority of sources will remain in the same frequency bin for the entire 
observation.  For these binaries, $t_i^\text{bin} =T_\text{obs}$, and the extra factor in 
\autoref{eq:GWB} becomes the standard $f_i^\text{obs}/\Delta f$.  

However, for the frequency range and binning that we consider, we cannot neglect evolving binaries 
despite their rarity.  For a binary to be shrinking quickly enough to change frequency bins over the 
course of the observation time, it must typically have a high mass and be very tight.  From 
\autoref{h1binary} it is clear that such binaries can produce particularly strong signals.  We 
therefore calculate the redshifted frequency for each source at a time $t + T_\text{obs}$.  For those 
binaries which have changed in frequency by at least one bin, we calculate the time spent in each 
bin, and divide up the signal accordingly. 

A single realization of the \textsc{gwb} for each simulation as a function of $\mchirp$ and $z$ is 
shown in \autoref{fig:hc_of_M_z}.  Although most binaries are at moderate mass and high 
redshift, most of the strain comes from binaries with $\mchirp \gtrsim 10^8 M_\odot$ and 
$0.1 \lesssim z \lesssim 2$.  Both simulations become incomplete for masses below 
$10^8 M_\odot$,  but the strain produced by low mass binaries primarily decreases due to the 
scaling of $h \sim \mchirp^{5/3}$, as can be seen at low redshifts where the bins are populated 
by individual binaries.  
On the other end of the scale, black holes with $\mchirp \sim 10^{10} M_\odot$ only 
rarely appear in individual realizations.  
Neither the increased inaccuracy of the scaling relations at low or very high black 
hole mass nor the missing low-mass binaries due to the simulation resolution should significantly 
affect the total strain.

\section{Discussion}
\label{sec:discussion}

In this section we present our calculated amplitude of the \textsc{gwb} and compare with other 
recent predictions. We discuss two forms of uncertainty on the amplitude of the \textsc{gwb}: that 
due to uncertainty in the astrophysical scaling relations and the variance of the \textsc{gwb} 
spectrum between individual realizations of the \textsc{smbbh} population.
Finally, we discuss the dependence of the realization-to-realization variance on the choice of 
frequency bins.

\subsection{Amplitude of the Characteristic Strain Spectrum}

\begin{figure}
	\includegraphics{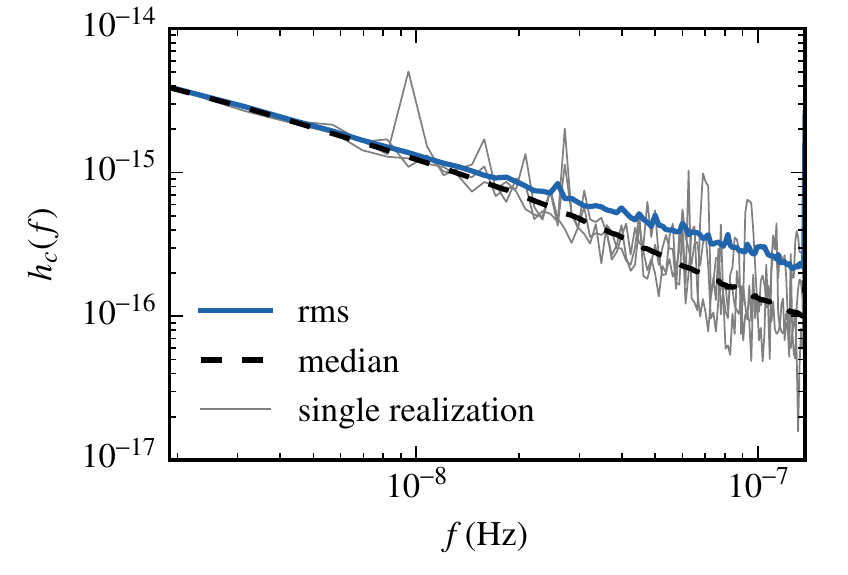}
	\caption{Three separate realizations of the \textsc{gwb} for MultiDark along with the median and rms
			 \textsc{gwb} calculated for the entire sample of 5,000 realizations.  At high frequencies,
			  the median signal drops below the canonical power law, and the amplitudes in 
			  adjacent bins become increasingly uncorrelated for non-evolving sources.}
	\label{fig:hc_realizations}
\end{figure}

The form of the \textsc{gwb} strain spectrum was calculated analytically by \citet{phinney01}, who 
determined that any sufficiently large isotropic population of compact objects in circular orbits decaying 
due to the emission of gravitational radiation should produce a stochastic background with a $2/3$ 
power law frequency spectrum.  In the real universe, the finite population of black holes will prevent 
the \textsc{gwb} from attaining power-law behavior, but a power law is expected to be approximately 
accurate for a subset of frequencies and represents the expected spectrum for an unobservable 
ensemble of realizations of the \textsc{gwb} \citep{sesana08}.  Pulsar timing arrays often search for 
a power law stochastic signal, and report constraints on the amplitude $A$ given by
\begin{equation}
	h_c(f) = A \left(\frac{f}{1 \text{ yr}^{-1}} \right)^{-2/3}.
\end{equation}
To date, the strongest upper bound on a power law frequency spectrum come from 
\citet{shannon15} who rule out a power law \textsc{gwb} with amplitude  $A > 1\times10^{-15}$
at 95\% confidence.  Other \textsc{pta}s \citep{arzoumanian15, lentati15} report similar constraints.

We generate 5,000 separate realizations of the \textsc{gwb} for each simulation, as described in 
\autoref{sec:gw_sig}.  Using the ensemble of spectra for each simulation, we calculate the rms 
and median strain.  An example is shown in \autoref{fig:hc_realizations}, where we plot three 
individual realizations of the spectrum, along with the rms and median strain.  As in \citet{sesana08}, 
at frequencies $\lesssim 10$~nHz the gravitational wave spectra are smooth and very similar to 
both the rms and median signals.  For frequencies $\gtrsim 20$~nHz, the amplitudes in each 
frequency bin vary widely and the median strain drops noticeably below the rms strain.

\begin{figure}
	\includegraphics{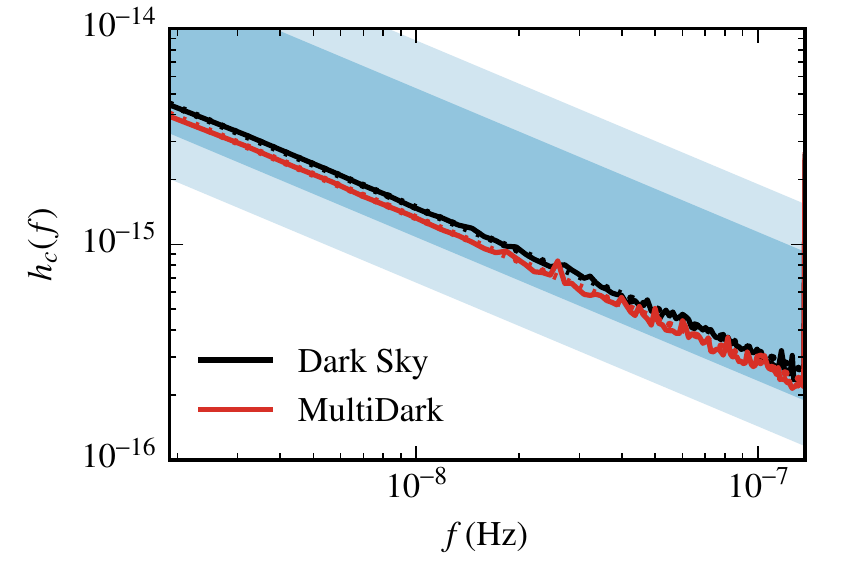}
	\caption{Rms amplitudes (solid lines) for the \textsc{gwb} in both simulations, as calculated 
			for 5,000 realizations assuming a 25 year observation time.  The dotted lines 
			represent best-fit $2/3$ power laws for Dark Sky (black) and MultiDark (red).  
			Shaded regions show $95\%$ confidence intervals on other recent predictions of 
			the \textsc{gwb}.  The larger pale blue region represents the \citet{sesana13a} 
			predictions when using the \citet{mcconnell13} scaling relations, which are most 
			comparable to our choice.  The smaller blue shaded region shows the \citet{ravi15a} 
			prediction.}
	\label{fig:sim_amplitudes}
\end{figure}

The rms spectrum (see \autoref{fig:sim_amplitudes}) is very similar to the expected $2/3$ 
power law, but with occasional spikes due to the presence of an extremely rare and bright source 
in that bin for one of the 5,000 realizations.  The best fit $2/3$ power laws have amplitudes 
$A_D = 6.84 \times 10^{-16}$ and $A_M = 6.14 \times 10^{-16}$. 

We also calculate the semi-analytic amplitude for each simulation by direct integration \citep{phinney01}:
\begin{align} 
	h_c^2(f) = \frac{4 G}{\pi c^2 f^2}\int_0^\infty \text{d}z \int_0^\infty \text{d}\mchirp \,
			\frac{\text{d}^2 n}{\text{d}z \, \text{d} \mchirp} \frac{1}{1+z} \frac{\text{d} E_\textsc{gw}}{\text{d} \ln f_\text{emit}},
\end{align}
where $n$ is the comoving number density and
\begin{align}
	\frac{\text{d} E_\textsc{gw}}{\text{d} \ln f_\text{emit}} = \frac{\pi}{3G} \frac{(G \mchirp)^{5/3}}{(\pi f_\text{emit})^{1/3}}.
\end{align}
The resulting semi-analytic amplitudes are $A_\text{D, SA} = 6.89 \times 10^{-16}$ and 
$A_\text{M, SA} = 6.13 \times 10^{-16}$, consistent with their respective Monte Carlo counterparts.  The slightly larger 
difference for Dark Sky is likely due to its coarser redshift resolution.

The difference in amplitude between the simulations is small but consistent with differences due to 
the initial halo mass functions and halo merger rates at low redshift. 
Our rms amplitudes are a factor of $\sim 2$ below the latest power law upper limits from \textsc{nanog}rav 
\citep{arzoumanian15}, and a factor of $\sim 1.5$ below the recent best upper limits from \textsc{ppta} 
\citep{shannon15}.  
Our rms amplitudes are consistent with the predictions of \citet{sesana13a} and \citet{ravi15a}, 
despite the different approaches used.

\subsection{Astrophysical Uncertainty}

\begin{figure}
	\includegraphics{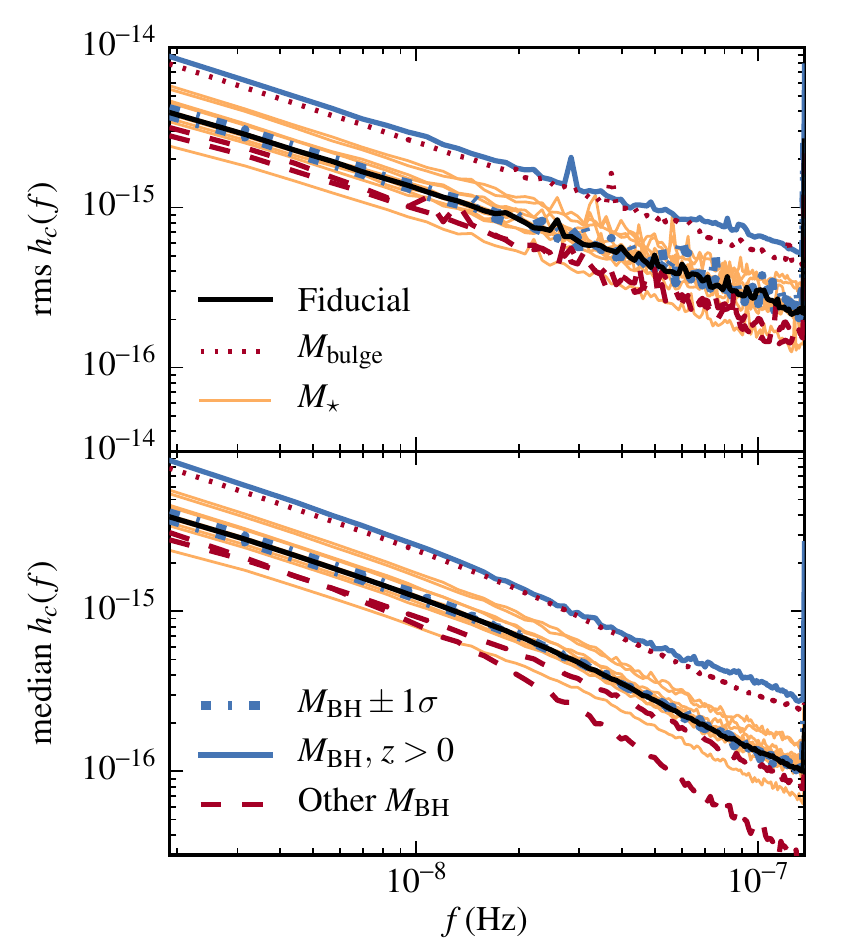}
	\caption{
		Rms and median strain of our fiducial MultiDark model (black).  Other lines show
		models with varying astrophysical scaling relations.  The single dotted 
		red line represents a model where bulge mass equals stellar mass for all galaxies.
		The thin yellow lines show 10 randomly selected points in the parameter space of 
		the \citet{behroozi13c} stellar mass--halo mass relations.  The dash-dotted blue lines 
		very close to the fiducial model are the $\pm 1 \sigma$ \citet{kormendy13} 
		$M_\text{bulge}$--$\mbh$ relations.  The dashed red lines represent the \citet{mcconnell13}
		(upper) and \citet{graham12} (lower) $M_\text{bulge}$--$\mbh$ relation.  The 
		solid blue line is the \citet{kormendy13} relation, but incorporating redshift 
		evolution as in \citet{bennert10a}.  
		}
	\label{fig:vary_astro}
\end{figure}

In this section, we briefly explore the effect of uncertainty in our astrophysical assumptions.  In 
particular, astrophysical uncertainties affect our calculations within the:
\begin{itemize}
	\item halo mass--stellar mass scaling relations,
	\item calculation of the bulge mass as a fraction of the total galaxy stellar mass,
	\item bulge mass--black hole mass scaling relation,
	\item galaxy merger rates, and
	\item processes used to form \textsc{smbbh}s and overcome the final parsec problem.
\end{itemize}
The effect of the final parsec problem and of the putative astrophysical processes which allow it 
to be overcome is a question of considerable scope.  We will leave its exploration for later work.
Additionally, we use the halo merger rates from the simulations to determine galaxy merger rates.  
Some uncertainty in the merger rate can be associated with the difference in amplitude between 
simulations, but we do not explore uncertainty in the galaxy merger rate further.
In what follows, we focus on the uncertainties due to the other three astrophysical processes listed 
above.  Results are shown in \autoref{fig:vary_astro}.

Our basic model for the galaxy stellar mass function uses the halo mass functions of the two 
simulations and the best-fit parameters of the \citet{behroozi13c} stellar mass--halo mass 
scaling relation.  To explore the uncertainty in this scaling relation, we repeat the calculation of 
the \textsc{gwb}, using 10 randomly chosen points from the allowed parameter space in the stellar 
mass--halo mass relation (P. Behroozi, private communication).  The results are shown as the 
thin yellow lines in \autoref{fig:vary_astro}.  Although most of the models have amplitudes very 
close to that of our basic model, outliers in both directions differ by a factor of $\sim 2$.

As discussed in \autoref{sec:place_bhs} our fiducial model calculates the bulge mass of 
galaxies by first assigning them to `quiescent' or `star-forming' populations and then using 
a population-dependent relationship between bulge mass and stellar mass.  This approach 
neglects the effects of environment and merger history.  To estimate the effect of our 
assumptions on the amplitude of the \textsc{gwb}, we use an alternate model in which the 
bulge mass equals the stellar mass for all galaxies, effectively assuming that all galaxies are 
elliptical.  This is not particularly plausible--for nearby galaxies with stellar masses 
$\gtrsim 10^{11} M_\odot$, elliptical and lenticular galaxies appear to make up roughly 
40\% of the total population \citep{hoyle12, wilman12}.  Instead, we use this as an upper limit on 
the amplitude that could be produced by varying this relation.  The resulting \textsc{gwb} 
is shown as a dotted red line in \autoref{fig:vary_astro}, and has an amplitude a factor of 2 
above the fiducial model.

To estimate the uncertainty on the \textsc{gwb} amplitude due to the black hole scaling relations, 
we repeat our calculations using the $\pm 1\sigma$ parameters from the \citet{kormendy13}
$M_\text{bulge}$--$\mbh$ scaling relations.  The resulting spectra are shown as dot-dashed blue
lines in \autoref{fig:vary_astro}.  They remain close to the original model.  
The two dashed red lines are calculated using the \citet{mcconnell13} and the \citet{graham12}
$M_\text{bulge}$--$\mbh$ scaling relations.  They are both within a factor of 2 of the amplitude 
in our fiducial model, although the \citet{graham12} broken power law model noticeably changes 
the shape of the median at high frequencies, and the rms spectrum is slower to converge to 
the expected shape.  This is not surprising given that this scaling relation produces relatively 
more low-mass binaries; see the discussion in \autoref{sec:cosmic_variance}.

We explore 
the possibility of evolution in the black hole scaling relation with redshift by applying the 
\citet{bennert10a} scaling of $\mbh/M_\text{bulge} \propto (1+z)^{1.4}.$  This produces 
an amplitude a factor of $\sim 2$ above our standard model, shown as the solid blue line in 
\autoref{fig:vary_astro}. 
However, this model might be better formulated to include a variant bulge mass model, since 
\citet{bennert10a} do not see any evolution with total galaxy luminosity, suggesting that 
bulge evolution plays a significant role in any evolution of the black hole scaling relation.

The primary effect of the different parameter choices for the varying relations is a vertical shift 
in amplitude in all bins; the shape of the spectrum is not strongly affected unless the black hole 
mass function changes significantly.  The range of amplitudes produced by astrophysical variants 
overlaps the \citet{ravi15a} and \citet{sesana13a} 95\% limits shown in \autoref{fig:sim_amplitudes}.  
The dominant effects appear in the stellar bulge mass calculation, but this may be modified by 
black holes growing at different eras from bulges.  Uncertainties in the stellar mass scaling 
relations and black hole scaling relations are smaller but of the same order of magnitude. 

The range in amplitudes suggested by our models for the astrophysical uncertainty is significantly
constrained by \textsc{pta} measurements.  The \textsc{ppta} upper limit on a power law \textsc{gwb} 
is within a factor of two of our fiducial amplitude, and rules out our variant models with either an 
evolving $\mbh$--$M_\text{bulge}$ relation or an all-elliptical galaxy population.  The \textsc{nanog}rav 
upper limit does not directly rule out any model presented here, but is in tension with the highest models.

\subsection{Variance from a Finite Number of Sources}
\label{sec:cosmic_variance}

In this section we discuss variance in the amplitude and shape of the \textsc{gwb} spectrum 
resulting from only having a single realization of the population of \textsc{smbbh}s.  In 
contrast to the effect of unknown astrophysics discussed in the previous section, this scatter 
is innate: it is the result of having only a single observation of the \textsc{gwb}, and is directly
analogous to the well-known effect of cosmic variance in the analysis of the cosmic microwave 
background power spectrum.  
It has been discussed previously in e.g.\ \citet{ravi12a, cornish13}, with earlier related work by 
\citet{jaffe03} and \citet{sesana08}.

The two forms of variance affect the spectrum of the \textsc{gwb} differently.  The astrophysical 
effects discussed in the previous section produce a systematic offset in the amplitude, while 
maintaining a power law spectrum.  In contrast, the variance due to only having a single 
realization does not produce a systematic offset in amplitude, but rather changes the spectrum 
away from a power law and adds an uncorrelated component to the amplitudes of the \textsc{gwb} 
in individual frequency bins.

In \autoref{fig:hc_realizations} we show three realizations of the \textsc{gwb} along with 
the rms and median strains calculated for the full set of 5,000 realizations.  Although the rms 
strain can be described by a power law, the shape of each realization cannot.
Each spectrum contains many spikes and dips due to the presence (or lack thereof) of 
bright individual sources \citep[cf.][]{rajagopal95, jaffe03, sesana08,kocsis11}.  At frequencies 
$\gtrsim 20$~nHz the rms signal is mostly due to these rare, high-amplitude events.  The median 
amplitude, as was originally found by \citet{sesana08}, dips below the rms signal.  At high 
frequencies, the scatter in the possible values of the amplitude increases, and the spectrum 
becomes very noisy.  

\begin{figure}
	\includegraphics{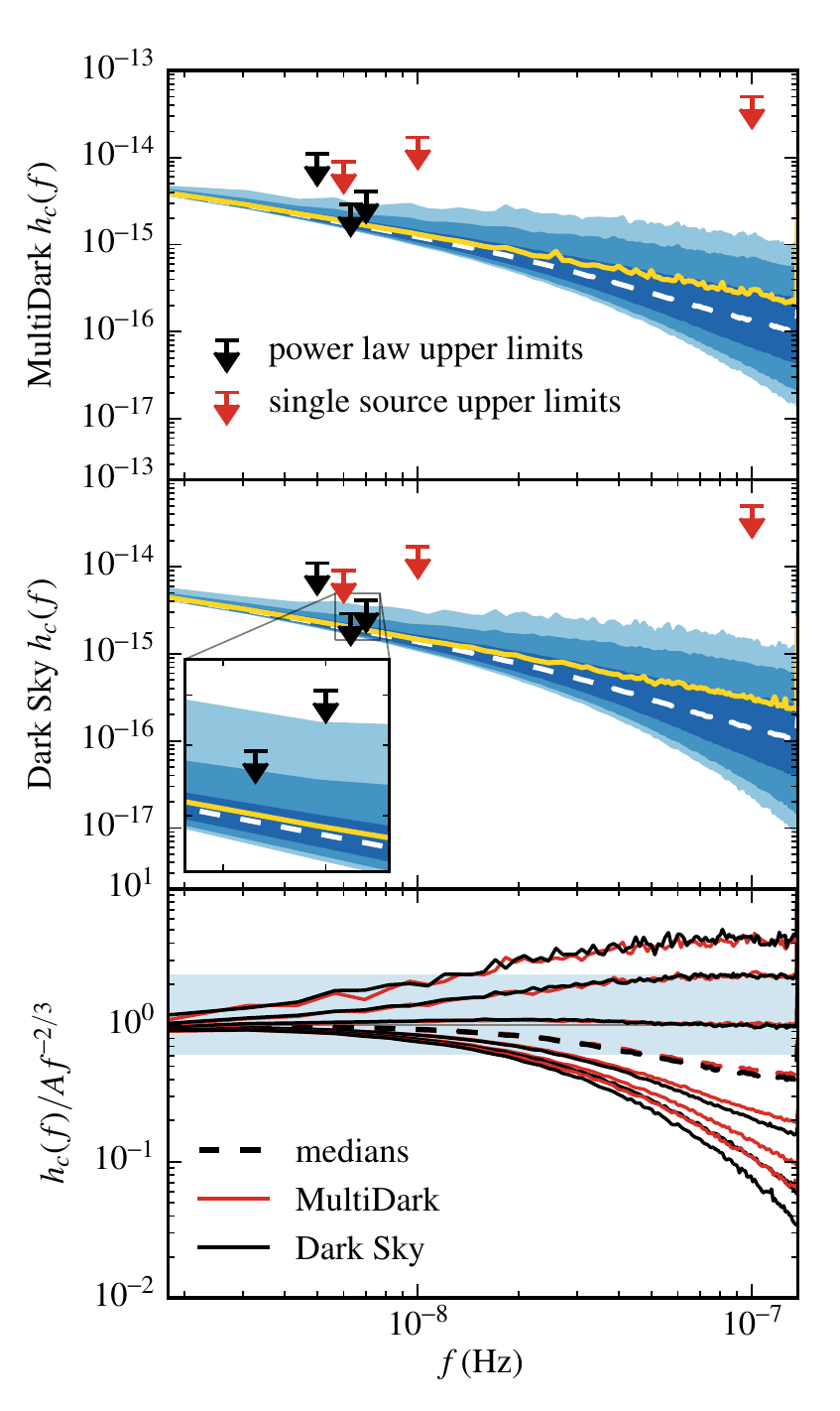}
	\caption{Scatter in the amplitude of the \textsc{gwb} spectrum calculated for 5,000 realizations 
			assuming a 25 year observation time.  Contours contain 68\%, 95\%, and 99\% of all 
			amplitudes in each frequency bin, measured from the median.  Top two panels: 
			white dashed lines show the median, and yellow solid lines show the rms amplitude.
			From left to right, the black upper limits represent the \textsc{epta} 
			\citep{lentati15}, \textsc{ppta} \citep{shannon15}, and \textsc{nanog}rav 
			\citep{arzoumanian15} upper 
			bounds on a $2/3$ power law \textsc{gwb}. The red upper limits represent the 
			\textsc{epta} \citep{babak16}, \citet{zhu14}, and \citet{yi14} upper limits on single 
			sources.  Bottom panel: the contours and medians shown above, but normalized by 
			the best fit power law for each simulation.  This is the cosmic variance of the 
			\textsc{gwb} around the underlying power law spectrum.  The pale blue band shows 
			the extent of the astrophysical uncertainty from \autoref{fig:vary_astro}.}
	\label{fig:amplitude_scatter}
\end{figure}

In \autoref{fig:amplitude_scatter} we plot the scatter in the amplitude of the gravitational wave 
spectrum in each frequency bin for all 5,000 realizations.  The scatter is represented by 
confidence intervals around the median amplitude for each frequency bin across our frequency 
range of interest.  In the lowest frequency bins, $99\%$ of all amplitudes differ by less than a 
factor of $2$, but by the highest frequency bins the spread between the ceiling and the floor of 
possible amplitudes has grown by more than an order of magnitude.  The third panel of 
\autoref{fig:amplitude_scatter} shows the deviation of the ensemble of observed spectra 
from the underlying power law spectrum, and represents the cosmic variance in the \textsc{gwb}.

The increased variance with frequency is primarily a result of \autoref{freq_decay}: heavy 
binaries have much shorter lifetimes than light binaries, and all binaries spend most of their lives 
at low frequencies.  As a result, the signal in the lowest bins is dominated by the largest binaries 
($\mchirp \sim 10^9 M_\odot$), but they pass through higher frequencies sufficiently quickly that 
they are unlikely to be seen in these bins.  Signals at higher frequencies are primarily due to lower 
mass binaries which have larger populations and live longer, but occasionally a high-mass binary 
will be observed.  These rare events have a significantly greater amplitude than is produced by 
the stable population of lower mass binaries, thereby increasing the scatter in the expected amplitude.  

Individual realizations of the \textsc{gwb} at low frequencies have a relatively smooth signal close to 
the $2/3$ power law, with few individually resolvable sources.  This is related to our choice of 
initial conditions for the binaries.  Binaries are assumed to have initially circular orbits, with initial 
frequencies consistently chosen outside the observation window.  These assumptions are required 
to derive a power law behavior as in \citet{phinney01}.  However, in the real universe, 
\textsc{smbbh}s will likely have some initial eccentricity before circularizing, and the mechanism used 
to solve the final parsec problem may move the black holes to higher frequencies than we have 
assumed.  Other astrophysical effects may also change the behavior of binaries at low frequencies, 
likely leading to a decrease in total signal.  Calculations that take these factors into consideration 
\citep[e.g.][]{enoki07, kocsis11, sesana13b, ravi14a} generally see a significant departure from 
power-law behavior at low frequencies.  

At higher frequencies the spectrum may vary significantly from bin to bin and will generally fall below 
the $2/3$ power law, but with some spikes rising above it.  Individually resolvable sources are more 
likely to be observed---note that the contour for the brightest $2.5\%$ of sources at $10^{-7}$~Hz is 
a factor of 2 higher than the rms power law and a factor of 5 above the median.

Additionally, evolving binaries become more frequent at high frequencies.  Binaries whose 
observed frequencies evolve through multiple bins over the course of the observation are rare 
events.  Since this can occur only near the end of the binary's lifetime, evolving binaries can be 
bright and represent a class of potentially resolvable events.  Sources that experience significant 
evolution produce a $2/3$ power law spectrum over the period of observations for the same reason 
as the overall population: although the amplitude of the signal increases as $f^{2/3}$, the time 
spent in each bin goes as $\sqrt{f^{-8/3}}$.  At frequencies $\gtrsim 50$~nHz for observations
$\gtrsim 10$~years, evolving binaries dominate the bright end of the space of possible amplitudes, 
and help ensure the power law behavior of the rms signal.  

In addition to searching for a power-law background, \textsc{pta} searches for individual 
(continuous wave) sources have been made \citep{jenet04, yardley10, arzoumanian14, zhu14, 
yi14, babak16}.  At low frequencies, the strongest constraints are due to \citet{babak16}, whose 
most robust $95\%$ upper limit on the strain amplitude is $h_c = 9\times 10^{-15}$ at a frequency of 
6~nHz.  This is within a factor of 2 of our $99\%$ upper limit and roughly a factor of 3 of our median 
or rms amplitude for both simulations.  At high frequencies, \citet{yi14} put upper limits on individual 
sources using high cadence observations of a single pulsar.  For randomly located sources, their 
upper limit on the strain amplitude is $h_c = 1.53 \times 10^{-11}$  at $10^{-7}$~Hz, which is 
4 orders of magnitude higher than our $99\%$ upper limits.  The constraint on optimally located 
sources improves to $h_c=4.99\times 10^{-14}$, although this remains more than an order of 
magnitude higher than our $99\%$ upper limits.

The probability distribution of the cosmic variance  within a single frequency bin is non-Gaussian, 
agreeing with the conclusions of \citet{ravi12a} and \citet{cornish13}.  This can be seen clearly in 
\autoref{fig:amplitude_scatter} and \autoref{fig:scatter_bins}, where a tail to high frequencies is 
readily apparent.  Indeed, from \autoref{eq:GWB}, the distribution of the power in each frequency 
bin is given by the sum of Poisson distributions, where each distribution is scaled by 
$h^2 \propto \mchirp^{10/3}$.   Sources drawn from the tails of these distributions will therefore 
dominate the signal in that frequency bin.  

As discussed in \citet{sesana08}, the exact shape of the median signal depends on the 
\textsc{smbbh} mass function.  Similarly, the amount of scatter in the amplitude at different 
frequency bins between possible realizations of the \textsc{gwb} will be sensitive to the relative 
quantities of \textsc{smbbh} of different masses.  In particular, the highest-mass binaries 
produce the strongest signal, but are extremely rare, so their relative abundance will affect the 
upper limits on the possible amplitudes of the signal for all frequency bins.  Smaller mass 
\textsc{bh}s are much more common and have much longer lifetimes, so there should be a 
population of them emitting in a wide range of frequencies in any single observation.  They 
produce vastly less signal ($h\propto \mchirp^{5/3}$) than higher-mass binaries, so the lower limits 
on the scatter in possible amplitudes will be set by the highest-mass population of binaries 
commonplace in each frequency bin.  
This is the reason for the significant increase in scatter as frequencies increase; at a few nHz, the 
very largest binaries have $\mchirp \sim 10^9 M_\odot$, but there are also enough binaries of this 
size that the highest-mass binaries which are common at that frequency are only slightly smaller.  
Near 100~nHz, the largest binaries have $\mchirp \gtrsim 10^8 M_\odot$ ($\mchirp \sim 10^9 M_\odot$ 
binaries are uncommon enough that they will rarely be seen at these frequencies), but the floor 
is produced by binaries with $\mchirp \sim 10^7 M_\odot$--$10^8 M_\odot$.  

The dependence of the floor on the largest commonplace population of binaries can be seen by 
comparing the lower bounds at high frequencies for the two simulations in \autoref{fig:amplitude_scatter}, 
along with the median in the \citet{graham12} model in \autoref{fig:vary_astro}. The latter two show a 
noticeable drop at frequencies of a few~$\times 10^{-8}$~Hz, whereas the fiducial MultiDark model shows a 
more gentle decline.  Although their mass functions are similar overall, MultiDark has more binaries at 
the lowest masses than Dark Sky. Similarly, the \citet{graham12} broken power law 
$\mbh$--$M_\text{bulge}$ model naturally leads to a smaller population of moderate mass binaries 
(and subsequently more low mass binaries) than our fiducial model.  
The missing binaries lead to a calculated \textsc{gwb} with bins at high frequencies where binaries of 
moderate masses that should be commonplace are instead relatively rare.  At these frequencies, the 
lower limits on the scatter in amplitude will decrease significantly.  
A similar, but milder effect is behind the downward turn of the $95\%$ and $99\%$ lower limits in MultiDark near
$3\times10^{-8}$~Hz.
This situation is unphysical: we are unable to accurately model binaries with 
$\mchirp < 10^8 M_\odot$, which are the binaries that we expect to define the floor of the scatter in 
the \textsc{gwb} at high frequencies.  We therefore expect less scatter to low strain than predicted by 
\autoref{fig:amplitude_scatter}.  

This argument only applies to the lower bounds, since the population producing the ceiling on the 
scatter in the \textsc{gwb} is well-defined. It is unlikely that even a significant population of missing 
lower-mass binaries would be able to affect the upper bounds; a binary at $z=0.1$ with 
$\mchirp=10^7 M_\odot$ would only add a strain of $h_c\sim10^{-17}$ to a frequency bin at 
$10^{-7}$~Hz.  Since \autoref{eq:GWB} varies as $\sqrt{n}$, it would take $\sim 100$ such 
missing binaries per frequency bin in every realization of the \textsc{gwb} to affect the median amplitude.  
A few missing binaries in each bin near $10^{-7}$~Hz in every realization, however, could bring the 
lower contours of Dark Sky into agreement with MultiDark.  The effect of smaller
binaries is even further reduced.  A similar binary with $\mchirp=10^6 M_\odot$ would produce a strain of 
$h_c\sim10^{-19}$, which means that $10^4$ such sources would be required to have the 
same effect as a single binary with $\mchirp=10^7 M_\odot$.

\subsection{Effect of bin size on scatter}
\label{sec:vary_bin_widths}

\begin{figure}
	\includegraphics{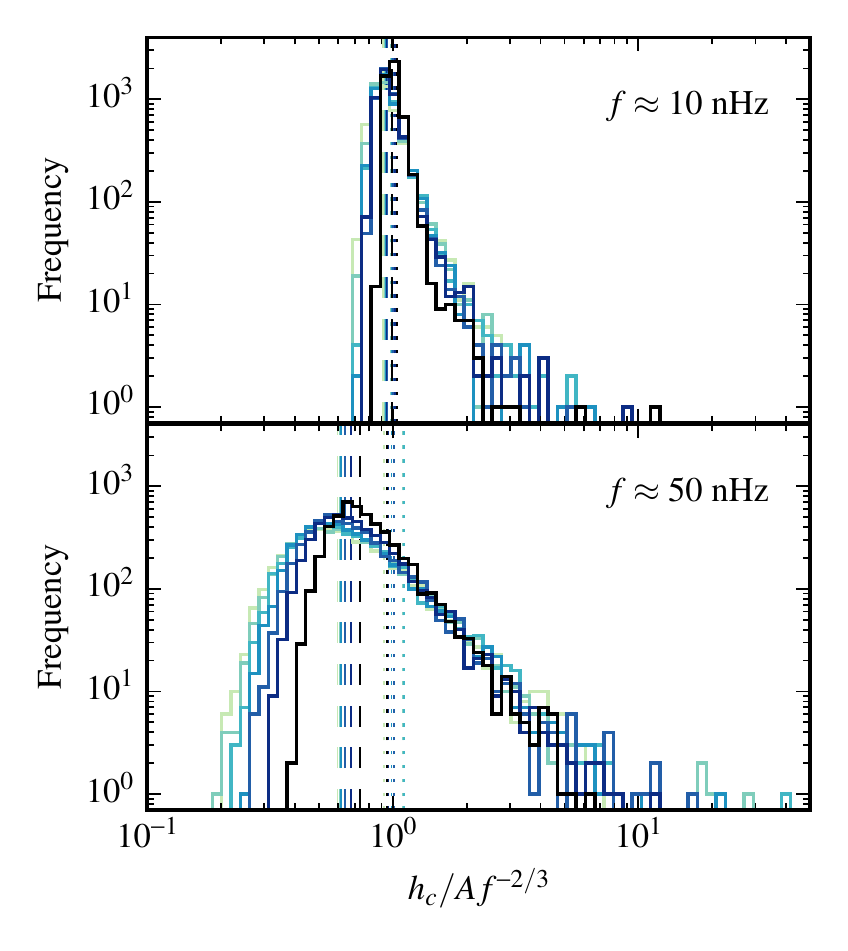}
	\caption{
	The cosmic variance shown at two different frequencies, calculated assuming different bin sizes.  
		From dark to light, histograms represent $\Delta f= 1/5$~yr, $1/10$~yr, $1/15$~yr, $1/25$~yr, $1/30$~yr, $1/35$~yr.
		Dashed lines show the median and dotted lines show the rms amplitudes for each histogram.}
	\label{fig:scatter_bins}
\end{figure}

 Thus far we have assumed a fixed set of frequency bins for all our calculations.  This is of little importance for the 
 calculation of the expected amplitude, but is not negligible when considering the scatter between realizations.  
 We repeat our calculations of the cosmic variance for MultiDark with a fiducial set of astrophysical parameters, but 
 assuming different total observation period $T_\text{obs}$, and therefore a different frequency bin width 
 $\Delta f = 1/T_\text{obs}$.  Results are shown in \autoref{fig:scatter_bins}.  
 
As $T_\text{obs}$ increases ($\Delta f$ decreases), the distributions broaden.   The broadening is most obvious 
to lower amplitudes (away from the tail), but the high-amplitude tails become broader as well.  This is unsurprising, 
given that the original source of this scatter is Poisson noise in the \textsc{smbbh} mass functions.  Narrower 
frequency bins will have fewer sources per bin, so the relative Poisson noise increases.  This is quite similar to the 
broadening of the distribution from low frequencies to high frequencies discussed in \autoref{sec:cosmic_variance}.
The rms amplitude remains approximately the same, since, as shown in \autoref{eq:GWB}, the characteristic amplitude 
is normalized by bin width, which is a proxy for the expected number of sources per bin.

Additionally, at frequencies near 100~nHz evolving sources begin to be important for spectra calculated with long 
observation times (narrow frequency bins).  The effect is to induce a correlation between neighboring bins.  Sources 
that move through many bins produce local $2/3$ power law spectra.  This in turn leads to a leveling of the cosmic 
variance contours, as can be seen in the third panel of \autoref{fig:amplitude_scatter} at frequencies $\gtrsim 70$~nHz.  
Since longer observation times and narrower frequency bins allow the evolution of more sources to be observed,
this effect increases with longer $T_\text{obs}$, but it remains unimportant at low frequencies for all observation 
times discussed.

\section{Conclusion}
\label{sec:conclusion}

We have used recent large-scale N-body simulations along with empirically calibrated scaling 
relations between dark matter halos and galaxies and between galaxy bulges and central black holes 
to calculate the expected \textsc{gwb} at low frequencies.  This approach is complementary to the recent 
calculations of the \textsc{gwb} based on empirical galaxy merger rates.  It also allows comparison with 
similar calculations based on the Millennium simulation, which is now over a decade old.  Our 
calculations of the \textsc{gwb} have only dealt with the sky-averaged signal, but 
can be readily extended to create mock sky maps encompassing direction and polarization 
information.  Such maps could be used to explore signal recovery by simulated \textsc{pta}s.  

We have calculated the amplitude of the canonical $2/3$ power law GW spectrum to be 
$\log A = -15.2^{+0.4}_{- 0.2}$, where the error represents the range of our variant astrophysical 
models and is not a $1\sigma$ Gaussian error.  
As shown in \autoref{fig:sim_amplitudes}, this range is consistent with amplitudes proposed by recent 
empirically-based calculations \citep{sesana13a, ravi15a}.  Our central amplitude is at the $3\sigma$ 
lower limits of the \citet{mcwilliams14} model, which assumes a high merger rate at low redshifts in order 
to reproduce the supermassive black hole mass function with mass growth through black hole mergers.  
Our model is also inconsistent with that of \citet{kulier15a}, 
which is calculated from hydrodynamical simulations of a galaxy cluster and a void.  These works make 
different assumptions about merger rates of galaxies, leading to the discrepant amplitudes and highlighting 
the strong dependence of the signal on astrophysical effects that are only beginning to be explored. 

We have estimated the uncertainty in the amplitude of the gravitational wave spectrum due to 
our incomplete understanding of the astrophysical effects involved.  In particular, we investigated
the halo mass--stellar mass relation, the $z=0$ and evolving black hole mass--stellar bulge mass, 
and the effect of a maximal model for the stellar bulge mass. The resulting range of 
amplitudes varies from our original model by a factor of $\sim 2$, and the shape of the spectrum 
remains essentially unchanged.  We also verified that the difference in amplitude produced by independent 
simulations assuming different recent cosmological parameters sets is much smaller than the variance 
due to astrophysical uncertainties.  This range in amplitude has already been constrained by 
the recent \textsc{ppta} \citep{shannon15} and  \textsc{nanog}rav results \citep{arzoumanian15}, ruling out 
power law amplitudes within a factor of two above our fiducial model.   

We have also characterized the scatter in amplitude expected for a single realization of the \textsc{gwb}.
This scatter represents a fundamental uncertainty in the amplitude expected in each frequency bin 
due to shot noise in the source population and the fact that only a single realization of the \textsc{gwb} can 
be observed.  As a result, the observed \textsc{gwb} will not follow a power law at all frequencies. At 
frequencies $\gtrsim 20$~nHz, the median amplitude becomes considerably lower than the 
power-law rms amplitude, with large scatter, suggesting that the \textsc{gwb} on average should be much 
harder to observe at these high frequencies.  However, rare bright sources also bring the signal in 
the corresponding frequency bins well above the $2/3$ power-law amplitude, so it might be possible 
to find individual sources at high frequencies.  

The two forms of uncertainty on the \textsc{gwb} investigated here---astrophysical and scatter between 
individual realizations---are fundamentally different.  Astrophysical uncertainty reflects our 
imperfect knowledge of the physics involved, and produces a systematic change in the 
amplitude of the \textsc{gwb}, without significantly affecting the underlying power-law shape of the
spectrum.  On the other hand, the scatter between individual realizations of the 
\textsc{gwb} is fundamental, much like cosmic variance in the study of the cosmic microwave 
background power spectrum.  Its presence will limit the degree to which the `true' underlying 
amplitude and shape of the \textsc{gwb} can be reconstructed, even with precise measurements.  
In contrast with the astrophysical uncertainty, the scatter between realizations is strongly frequency 
dependent.  As a result, we can expect that the dominant uncertainty at frequencies $\lesssim10$~nHz 
is astrophysical, and any single realization of the spectrum should look close to a power law at these
frequencies, although this becomes less true as observation times lengthen and frequency bins narrow.  

One class of astrophysical uncertainties retains the potential to muddy this distinction.  Mechanisms 
that allow the final parsec problem to be solved may continue to produce environmental effects on 
binaries that are in the process of emitting gravitational radiation. This affects the rate at which the 
orbits decay, typically suppressing the low-frequency amplitude of the \textsc{gwb} \citep{kocsis11, 
sesana13b, ravi14a}.  Such environmental effects change the number of binaries common at a given 
separation, as would the presence of stalled binaries, potentially affecting the scatter in the amplitude, 
and depressing the low-frequency end of the \textsc{gwb} below the nearly power-law behavior shown in 
\autoref{fig:amplitude_scatter}.  As previously shown by e.g.\ \citet{ravi14a, enoki07}, another 
possible environmental effect is increased eccentricity at wide separations, which suppresses the 
low-frequency \textsc{gwb} since eccentric binaries emit gravitational radiation in a series of harmonics 
rather than at a single frequency.

Such effects would not only affect the amplitude of the \textsc{gwb}, but would also change its 
underlying form away from a power law, and would likely increase the scatter between individual 
realizations at low frequencies.  In this case, the separation of the spectrum into a power law 
frequency band, where astrophysical uncertainties dominate, and a shot noise-dominated band 
may be complicated.  There are three potential scenarios:
\begin{itemize}
\item The first is our default assumption in this paper and that made by most \textsc{pta} searches: all 
energy and angular momentum loss is due to gravitational wave emission and binaries are circular.  
The spectrum will separate into power law and shot noise-dominated regions.  
\item The second is of particular interest in studying the final parsec problem.  In this scenario, binaries 
may be elliptical and the binary may lose energy and angular momentum due to processes other 
than gravitational wave emission, but only at the lowest frequencies measured. The spectrum 
would then be composed of  non-power law behavior or increased scatter at the lowest 
frequencies, the standard power law at moderate frequencies, and shot noise at the highest 
frequencies.  
\item In the third scenario, the effects of ellipticity or other processes
  continue well into band, resulting in the absence of a clear power law region.
  This scenario would likely produce spectra that are more  
difficult to measure and analyze, since they are likely to be noisy and have a low amplitude at 
all frequencies.  
\end{itemize}

Once gravitational radiation in the nHz regime has been detected, measuring its spectrum will 
be of much interest. 
We find that the effects of Poisson statistics dominate the
  uncertainties at the high frequency ($\gtrsim 20$~nHz) regime, while the astrophysical
  uncertainties dominate at lower frequencies.
The solution to the final parsec problem will be encoded into the shape 
of the spectrum of the stochastic \textsc{gwb}, but the recovery of its underlying form from 
the measured spectrum may require characterization of the astrophysical uncertainties and 
an understanding of the effects of cosmic variance in the gravitational wave background.

\acknowledgments{
We thank Alberto Sesana for insightful comments and Peter Behroozi for the use of a portion of the \textsc{mcmc} 
chain used to fit parameters in the stellar mass--halo mass relation.  
Computations were made on the Guillimin supercomputer from McGill University, managed by 
Calcul Qu\'ebec and Compute Canada. The operation of this supercomputer is funded by the 
Canada Foundation for Innovation (\textsc{cfi}), NanoQu\'ebec, \textsc{rmga}, and the Fonds 
de recherche du Qu\'ebec--Nature et technologies (\textsc{frq-nt}).
GPH acknowledges support from the \textsc{nserc} Discovery program, the Canadian Institute for 
Advanced Research, and the Canada Research Chairs program.
DEH was supported by \textsc{nsf career} grant PHY-1151836. He also acknowledges support 
from the Kavli Institute for Cosmological Physics at the University of Chicago through \textsc{nsf}
grant PHY-1125897 as well as an endowment from the Kavli Foundation.
}

\bibliography{gw_background}

\end{document}